\documentclass[aps,prl,showpacs,floatfix,twocolumn,longbibliography]{revtex4-1}
\usepackage{mathrsfs}
\usepackage[figuresright]{rotating}
\usepackage{amsmath,upgreek}
\usepackage{amssymb}
\usepackage{graphicx}
\usepackage{color}
\usepackage{dcolumn}
\usepackage{bm}
\usepackage[breaklinks=true,colorlinks=true,linkcolor=blue,urlcolor=blue,citecolor=blue]{hyperref}
\usepackage{ulem}
\usepackage{soul}
\usepackage{natbib}

\def\sro{Sr$_2$RuO$_4$}

\def\gyro{$^{17}\gamma$}

\def\cc{\mathbf{c}}
\def\BB{\mathbf{B}}

\def\be{\begin{equation}}
\def\ee{\end{equation}}

\def\deg{^{\circ}}

\makeatletter

\newcommand{\Rmnum}[1]{\expandafter\@slowromancap\romannumeral #1@}
\makeatother

\begin{document}

\title{Evidence for even parity unconventional superconductivity in \sro}

\author{A. Chronister$^{1 \dagger}$}
\thanks{These authors contributed equally}
\author{A. Pustogow$^{1 \dagger}$}
\thanks{These authors contributed equally}
\author{N. Kikugawa$^{2}$}
\author{D. A. Sokolov$^{3}$}
\author{F. Jerzembeck$^3$}
\author{C. W. Hicks$^{3}$}
\author{A. P. Mackenzie$^{3,4}$}
\author{E. D. Bauer$^{5}$}
\author{S. E. Brown$^{1}$}
\email[]{Address correspondence to: aaronchronister@ucla.edu, pustogow@physics.ucla.edu, brown@physics.ucla.edu.}
\address{$^1$Department of Physics $\&$ Astronomy, UCLA, Los Angeles, CA 90095, USA;}
\address{$^2$National Institute for Materials Science, Tsukuba 305-0003, Japan;}
\address{$^3$Max Planck Institute for Chemical Physics of Solids, Dresden 01187, Germany;}
\address{$^4$Scottish Universities Physics Alliance, School of Physics and Astronomy, University of St Andrews, North Haugh, St Andrews KY16 9SS, UK;}
\address{$^5$Los Alamos National Laboratory, Los Alamos, New Mexico 87545, USA;}

\date{\today}

\begin{abstract}

Unambiguous identification of the superconducting order parameter symmetry of \sro\ has remained elusive for more than a quarter century. While a chiral $p$-wave ground state analogue to superfluid $^3$He-$A$ was ruled out only very recently, other proposed $p$-wave scenarios are still viable. Here, field-dependent $^{17}$O Knight shift measurements 
are compared to corresponding specific heat measurements, previously reported. We conclude that the shift results can be accounted for by the expected field-induced quasiparticle response only. An upper bound for the condensate magnetic response of $<10\%$ of the normal state susceptibility is sufficient to exclude odd-parity candidates.
\end{abstract}
\maketitle

%

Unraveling the secrets of the superconducting state in \sro~\cite{Mackenzie2017,Kallin2012,Mackenzie2003} has been a priority for unconventional superconductivity research since its discovery in 1994 by Maeno and coworkers~\cite{Maeno1994}. Particularly notable among several reasons for broad interest in \sro\ was the suggestion of a $p$-wave triplet pairing state~\cite{Rice1995}. Among states allowed by symmetry is the chiral state $\mathbf{z}(p_x\pm ip_y)$, which breaks time reversal symmetry and therefore requires two components. Soon after, the combination of results from NMR Knight shift \cite{Ishida1998} and $\mu^+$SR \cite{Luke1998} measurements lent support to the chiral $p$-wave description. Further evidence was inferred from the observed onset of a non-zero Kerr rotation at $T_c$~\cite{Xia2006}. Unresolved issues remained, however. For example, thermal conductivity \cite{Hassinger2017} and specific heat \cite{Kittaka2018} experiments were both interpreted as evidence for a nodal gap structure \cite{Mackenzie2003}. In a step toward clarification, recent $^{17}$O NMR measurements excluded any candidate $p$-wave state with a $\mathbf{d}$-vector aligned parallel to the $c$-axis \cite{Pustogow2019,Ishida2020}. The NMR results published so far leave open the possibility for states with an in-plane ${\mathbf{d}}$, as has since been explicitly discussed in several theoretical works~\cite{Romer2019,Roising2019,Lindquist2019}. 

With these developments in mind, we recall other physical properties of \sro\ deemed relevant to the order-parameter symmetry. Recognized early \cite{Maeno1994,Rice1995} was the evidence for ferromagnetic correlations, as originally inferred from the measured Wilson ratio exceeding unity. Meanwhile, quantum oscillations \cite{Mackenzie1996}, ARPES \cite{Damascelli2000,Tamai2019}, and optical conductivity experiments~\cite{Stricker2014} all indicate modest normal-state mass enhancement. At the same time, an unconventional ground state was indicated by its very strong suppression with non-magnetic impurities~\cite{Mackenzie1998}. Thus, \sro\ constitutes a model system for the emergence of an unconventional superconducting state at low temperature, from a strongly correlated and ultraclean Fermi-liquid normal state.

The temperature and field dependences of the NMR Knight shifts $K_s(T<T_c,\mathbf{B})$ are a crucial probe of the 
~order-parameter symmetry, in particular for \sro. In the normal state, $K_s\sim\chi_n$, with $\chi_n$ the susceptibility. In the superconducting phase, a nonzero susceptibility $\chi_{sc}$ associated with condensate polarization is expected generally for triplet-paired, $p$-wave states.
Hence, the observed reduction of the Knight shift for an applied in-plane field excludes the chiral state~\cite{Pustogow2019}, for which $\mathbf{d}\parallel\mathbf{c}$. Not eliminated are states characterized by $\mathbf{d}\perp\mathbf{c}$. 

At first sight, the most direct way to test for symmetry-allowed states with $\mathbf{d}\perp\cc$ is to perform measurements with $\BB\parallel\cc$.  However, the relevant upper critical field $B_{c2,[001]}<1$ kG is very small~\footnote{$a$-axis stress increases $B_{c2}$ significantly by this measure, see Ref.~\cite{Steppke2017}} making such experiments particularly challenging because signal strength and spectral resolution are reduced for very weak applied fields. Here, we take another approach, discussed previously in Refs.~\cite{Ishida2020,Amano2015}: the field orientation is fixed in-plane, and the $^{17}$O shifts $K_s$ are evaluated at low temperature (25 mK) while varying $B$ as much as experimentally feasible. Quasiparticle creation is controlled by the field strength, and also contributes to the magnetic response. By way of comparing to previously reported specific heat results $C_e(B)/T$~\cite{Kittaka2018,NishiZaki2000}, we estimate the upper bound for the condensate portion to $\chi_{sc}/\chi_n<10\%$, a value that contradicts the expectation for any of the pure $p$-wave order parameters relevant for Sr$_2$RuO$_4$.

\begin{figure}
\includegraphics[width=1\columnwidth]{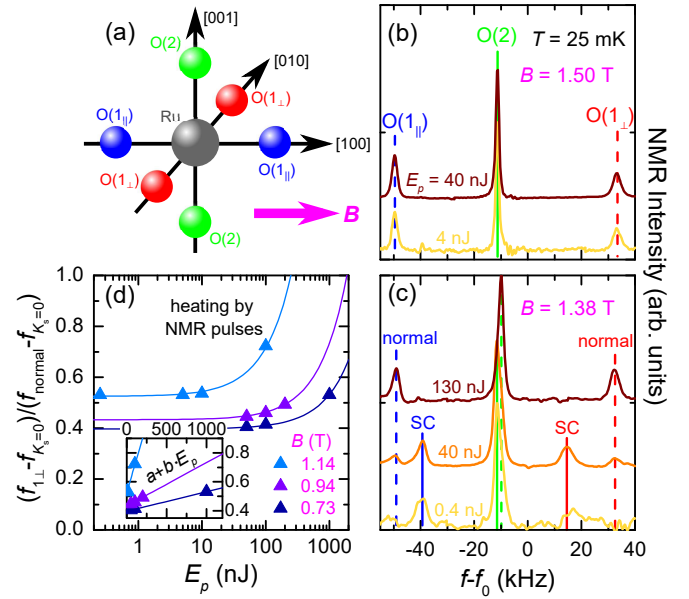}
\caption{(a) \sro\ involves three distinct oxygen sites for field direction $\mathbf{B}\parallel [100]$.
(b) The three associated $^{17}$O NMR central transitions (O(1$_{\parallel}$), O(2), O(1$_{\perp}$) from left to right) are independent of pulse energy $E_p$ at 1.50~T $>B_{c2}\simeq 1.45$~T. (c) Also at $B = 1.38\textrm{T}\lesssim B_{c2}$ the normal-state spectrum is observed for $E_p\geq 10^{-7}$~J. Reducing to $E_p=40$ nJ leads to doubled spectral features, most pronounced for O(1$_{\parallel,\perp}$), which we assign to coexisting normal (dashed vertical lines) and superconducting (solid) contributions around the first-order transition. Further reduction of $E_p$ reveals the pure superconducting-state spectrum. (d)  O(1$_{\perp}$) frequencies normalized to normal-state ($f_{normal}$) and zero-shift ($f_{K_s=0}$; see Fig.~\ref{fig:FieldSpectra}) positions at $B<B_{c2}$ for variable $E_p$. Linear fits (solid lines, see inset) indicate that heating is less problematic at lower field due to larger $T_c(B)$. Knight shifts $K_s$ were determined using the frequency values leveling off at $E_p\rightarrow 0$.
}
\label{fig:SROwarm}
\end{figure}

As in previous NMR studies on \sro~\cite{Ishida1998}, the labelled $^{17}$O ($^{17}I$=5/2, \gyro=-5.772 MHz/T~\cite{Harris2001}) is introduced by high-temperature annealing~\cite{Ishida1998}, here in 90\% $^{17}$O$_2$ atmosphere at 1050 $^\circ$C. To facilitate access to relatively low frequencies covering several octaves, we adopted a top tuning/matching configuration. 
Single-crystal dimensions were (3.5 mm x 1 mm x 0.2 mm), with the shortest dimension corresponding to the out-of-plane [001]-direction, and the longest dimension parallel to [100], see Fig.~\ref{fig:SROwarm}(a). The NMR coil containing the crystal under study, was mounted on a single-axis piezo-rotator inside the mixing chamber of a bottom-loading dilution refrigerator. Sample alignment enabled in-plane orientation to within $\pm0.2^\circ$, based on RF susceptibility measurements sensitive to $B_{c2}$, described in Ref.~\cite{Pustogow2019}, and discussed in the Supplemental Material~\cite{SM}. 
~$^{63}$Cu NMR relaxation rate measurements were used to determine the equilibrium bath temperature $T=25$ mK. As in our previous work~\cite{Pustogow2019}, low-power RF experiments were carried out to make sure the results were not measurably altered by RF pulse heating effects. The applied field strength $B$ was determined to within uncertainties less than 10's of $\mu$T from the NMR resonance of $^3$He in the $^3$He/$^4$He mixture of the dilution refrigerator.

Addressed first are sample heating effects by the RF pulses, illustrated in Fig.~\ref{fig:SROwarm}, which turned out to be a crucial issue~\cite{Pustogow2019,Ishida2020}. So as to enhance sensitivity to this potential artifact, we examined the transients with the field set to 1.38 T, a value just smaller than $B_{c2}$. Clear evidence for warming by the RF pulsing is inferred from a transient response corresponding to that of the normal-state (instead of the sought-after superconducting state). Shown in Fig.~\ref{fig:SROwarm}(b,c) are $^{17}$O spectra corresponding to central transitions for the three sites, O(1$_{\parallel}$,2,1$_{\perp}$), at applied magnetic fields slightly above and below $B_{c2}$. While at 1.5~T $>B_{c2}$ the line shape remains unaffected by changing the pulse energy, a normal state spectrum is produced also at 1.38 T $<B_{c2}$ when using a pulse energy $E_p$ = 130 nJ. Decreasing $E_p$ to 40~nJ leads to a response where a new spectral line appears for each site, indicating the coexistence of normal and superconducting phases. This data set is particularly useful, since the macroscopic phase segregation provides a quantitative measure of the magnetization jump $\Delta M$ at the discontinuous (first-order) transition ~\cite{Yonezawa2013,Yonezawa2014}. Note that these data are recorded following a single-pulse excitation. That is, the transient NMR response corresponds to a free induction decay (FID). All shift results of the present work were obtained from FID measurements carried out with RF pulse energies low enough to avoid heating, as illustrated in Fig.~\ref{fig:SROwarm}(d).

\begin{figure}[ptb]
\centering
\includegraphics[width=1\columnwidth]{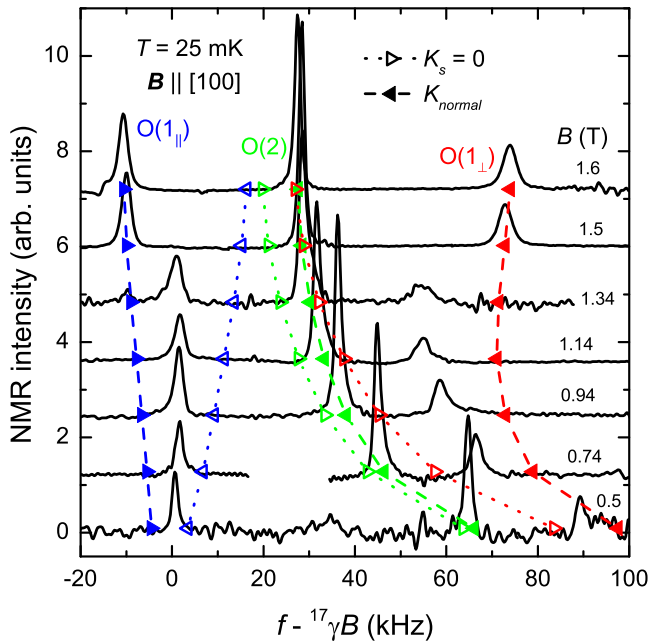}
\caption{Spectra for central $^{17}$O NMR transitions at different field strengths, for O(1$_{\parallel}$), O(2), O(1$_{\perp}$) sites, respectively \textit{left}-\textit{right}, plotted as intensity \textit{vs.} $f-$%
$^{17}\gamma B$. The dotted curves running vertically through the spectra follow the expected field dependence after taking into account quadrupolar and orbital couplings; the dashed curves also include the normal-state hyperfine fields. See Ref.~\onlinecite{SM} for details of quadrupolar and orbital contributions to the transition frequencies, as well as an analysis of the sample orientation relative to $\mathbf{B}$. }\label{fig:FieldSpectra}
\end{figure}

Having established a threshold for heating effects, we now inspect the spectra recorded at variable field strength. In Fig.~\ref{fig:FieldSpectra} we show the NMR intensity as a function of $f-f_0$, where $f_0\equiv^{17}$%
$\gamma B$. The central transitions ($-1/2\longleftrightarrow1/2$) for the O(1$_{\parallel}$,2,1$_{\perp}$) sites [left to right in the spectrum] exhibit pronounced variations with changing $B$. The shifts of the planar sites O(1$_{\parallel}$) and O(1$_{\perp}$) have opposite sign; this is a consequence of the applied field direction relative to the local environment. O(2) is the apical site [Fig.~\ref{fig:SROwarm}(a)]. 
The dotted curves include only the quadrupolar and orbital contributions for each site, while omitting the Knight shift contribution; more information on these corrections appear below and in Ref.~\onlinecite{SM}. Open symbols line up with these spectral ``baselines'' at each field at which data were recorded. Also shown, using the dashed lines and closed symbols, are transition frequencies at each field, generated using the \textit{known} normal state NMR parameters~\cite{SM}. Then, the frequency \textit{differences} between closed and open symbols are proportional to the hyperfine fields, and constitute the product of (normal-state) Knight shifts with applied field, $K_{s,\textrm{normal}}$%
$^{17}\gamma B$, for O(1$_{\parallel}$), O(2) and O(1$_{\perp}$). When decreasing the field $B<B_{c2}$, the NMR lines in Fig.~\ref{fig:FieldSpectra} are displaced from the normal-state positions, towards the frequency corresponding to $K_s=0$, due to the drop of $K_s$ in the superconducting state.
Below, we compare and contrast the measured shifts $K_s$ with results of field-dependent specific heat experiments, which are sensitive to the field-induced quasiparticles.

The parameters needed to make the quadrupolar corrections were determined previously~\cite{Imai1998,Ishida1998,Luo2019} and confirmed here in field-dependent measurements~\cite{SM}. In particular, we determined the field orientation as deviating $3^{\circ}$ from the [100] direction, and otherwise aligned orthogonal to the $c$-axis, $\theta=90^\circ\pm0.2^\circ$. Due to several factors, including reduced signal strength and resolution, as well as the strong increase of the O(1$_{\parallel}$) quadrupolar component at low fields, we limited the measurements to $B\geq 0.24$ T. In addition to the well-known quadrupolar effects, one has to include purely orbital contributions.
~These were evaluated in Ref.~\cite{Ishida1998}, yielding $K_o=+0.18\textrm{\%}$ for the O(1$_{\parallel}$) site and a value indistinguishable from zero for O(1$_{\perp}$) and O(2). See Ref.~\onlinecite{SM} for further comment.

\begin{figure}[tb]
\centering
\includegraphics[width=1\columnwidth]{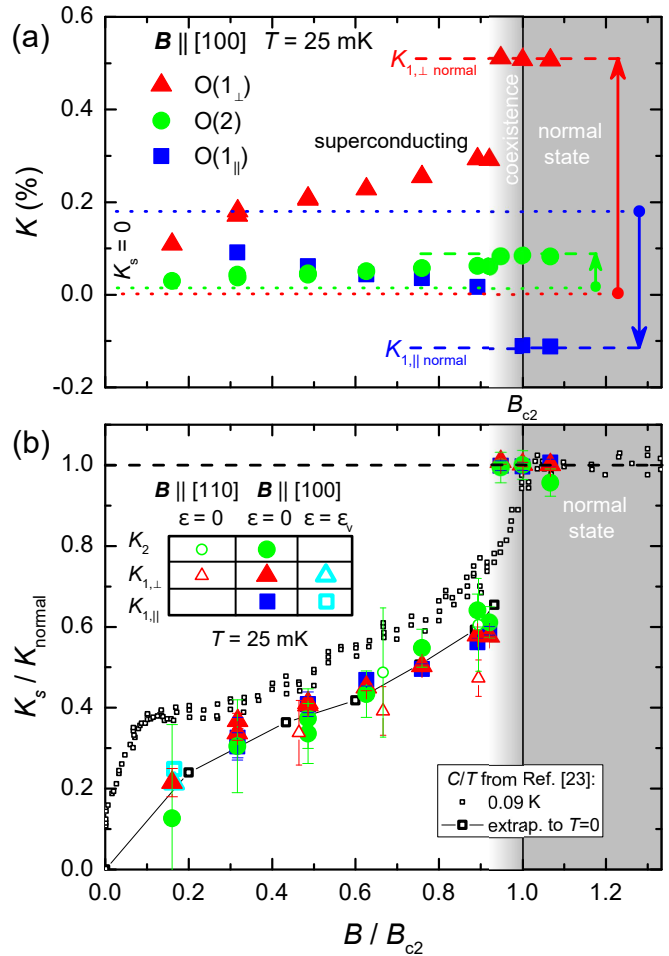}
\caption{(a) NMR shifts $K=K_s+K_o$ determined from the spectra in Fig.~\ref{fig:FieldSpectra}. While the shifts are positive and the assigned $K_o\simeq0.0\%$ for O(2) and O(1$_{\perp}$), the O(1$_{\parallel}$) line occurs at a positive value $K_o=0.18\%$ at $B=0$ and $K_{1,\parallel}<0$~\cite{Imai1998,Ishida1998}.
(b) The field-dependent drop of NMR Knight shift determined in the present work at $T=25$~mK is compared to specific heat $C/T$ recorded at $T=90$~mK~\cite{NishiZaki2000} as well as its $T=0$ extrapolation \cite{CpComment}, all normalized to the normal state value.
~The values of $K_s$ coincide with the zero-temperature extrapolations of $C/T$
, providing compelling evidence that this is the contribution of unpaired quasiparticles in the superconducting state. Measurements along [110] (small open symbols) reveal a similar jump at the transition and also uniaxial strain results (open cyan symbols, $\BB\parallel[100]$, $\varepsilon_{aa}=\varepsilon_{v}$) from Ref.~\onlinecite{Pustogow2019} coincide at low $B/B_{c2}$. 
}
\label{K_C_kappa}
\end{figure}
The shifts $K_{1\parallel,2,1\perp}$, are plotted as a function of $B$ in Fig.~\ref{K_C_kappa}. Results are shown in panel (a) as total shift, $K=K_s+K_o$. In the normal state, $K_{1\parallel}<0$, while $K_{2,1\perp}>0$; each exhibits a reduction in the superconducting state. $B_{c2}$ is marked by the discontinuous change of each of the three sites, accompanied by a coexistence regime [cf. Fig.~\ref{fig:SROwarm}(b,c)]. 
~Consistent with expectations ($B\gg B_{c1}$)~\cite{Murakawa2007}, the results indicate that diamagnetic shielding is a small effect. Otherwise, the discontinuous drop $\Delta M$ (Figs.~\ref{fig:SROwarm},\ref{fig:FieldSpectra}) would be similar for all three sites. Instead, only the hyperfine field, which is much greater for the planar sites than it is for the apical site, and opposite in sign for O(1$_{\parallel}$) relative to O(2) and O(1$_{\perp}$), decreases on entering the superconducting state.

The main results of this work are displayed in Fig.~\ref{K_C_kappa}(b), where the Knight shifts are compared to previous heat capacity results~\cite{NishiZaki2000}, $C_e(B)/T$ ($C_e$ the electronic contribution), both normalized to the normal state. As shown, the field-induced trends are similar, and particularly relevant to the open question of order-parameter symmetry. Simply put, at non-zero field, a Zeeman-like NMR response can originate from quasiparticles, and, in the case of triplet pairing, also from the condensate. In contrast, the specific heat is sensitive only to the quasiparticle response with no contribution from the condensate.
As can be seen by inspection of Fig.~\ref{K_C_kappa}(b), we observe no systematic difference between the $T\to0$ extrapolation of the heat capacity data of Ref.~\cite{NishiZaki2000} and the spin susceptibility deduced from our measurements.  Taking into account systematic uncertainties primarily associated with the oxygen orbital shifts $K_o$, we estimate an upper limit for the condensate response of $<10\%$ of that of the normal state, for fields applied both along [100] and [110] (see Ref.~\onlinecite{SM} for comments pertaining to $K_o$). Similar $K_{1\parallel,\perp}$ are found at $B/B_{c2}=0.17$ under strained conditions~\cite{Pustogow2019}, where $K_o$ is less of an issue due to larger field resulting from the enhanced $T_c$ and $B_{c2}$~\cite{Steppke2017}. These observations place such strong constraints on the magnetic polarizability of the condensate that we believe that they rule out any pure $p$-wave order parameter for the superconducting state of \sro, as we now discuss.

The $p$-wave order parameters most commonly discussed in the context of \sro\ are the so-called chiral ($\hat{\mathbf{z}}(p_x\pm ip_y)$) and helical ($p_x\hat{\mathbf{x}}+p_y\hat{\mathbf{y}}$) states.  Assuming that the unit vectors encoding spin directions are pinned to the lattice, they are predicted in the simplest models to result in condensate polarizabilities of 100\% (chiral) and 50\% (helical) of the normal state value.  The chiral state was ruled out by our previous work \cite{Pustogow2019}, but the helical state and certain others were not.  The data presented in Fig.~\ref{K_C_kappa} allow us to go much further.  Even after considering Fermi-liquid corrections~\cite{Ishida2020} and the effects of spin-orbit coupling~\cite{Roising2019}, it is unclear how to reconcile an upper limit of 10\% of the normal state susceptibility with any $p$-wave state. One could also postulate extreme situations such as a momentum independent $\mathbf{d}$ aligned along either [100] or [110], or an unpinned $\mathbf{d}$ free to rotate in response to the applied field.  None can predict a spin susceptibility suppression that would be compatible with our results; a few remaining possibilities have been ruled out by our use of both [100] and [110] fields in the current experiments.
We therefore assert that our measurements have ruled out any pure $p$-wave order parameter candidate for the superconducting state of \sro.

Given this input, we close with an evaluation of the current understanding of superconductivity in \sro. In isolation, our NMR findings are consistent with even-parity states, such as $d_{x^2-y^2}$ ($B_{1g}$), $d_{xy}$ ($B_{2g}$) or $\{d_{xz};d_{yz}\}$ ($E_{1g}$). Indeed, STM measurements are interpreted as most consistent with the $B_{1g}$ state~\cite{Sharma2020}, similar to thermal transport experiments~\cite{Hassinger2017}. Accordingly, $C_e/T$ follows the expected $\sqrt{B}$ dependence~\cite{Matsuda2006} in the zero-temperature limit.

However, several other experiments must be accounted for. These include recent ultrasound reports of a discontinuity in the elastic constant $c_{66}$~\cite{Ghosh2020,Benhabib2020}, which would restrict possible states to two-component candidates, such as $\{d_{xz};d_{yz}\}$. Time-reversal symmetry breaking may prove crucial~\cite{Grinenko2020}. Following the two-component hypothesis, while avoiding a reliance on interplanar pairing, has led to considering the possibility for coupling between accidentally nearly degenerate single-component states such as $\{d_{x^2-y^2};g_{xy(x^2-y^2)}\}$~\cite{Zutic2005,Suh2019,Kivelson2020}. It will be intriguing to see how the quest to finalize identification of the order parameter of \sro\ develops.  We believe that by ruling out any pure $p$-wave order parameter possibility, the research we have reported here makes a significant contribution to that process.


Note added: We recently learned of a proposal~\cite{Scaffidi2020} for a mixed-parity order parameter of the form $d\pm ip$~\cite{Scaffidi2020}, which would result in a reduced condensate response, relative to those of the the pure $p$-wave states discussed above. For example, $K_s/K_n\simeq0.2$ (chiral $p$-wave component), $K_s/K_n\simeq0.1$ (helical $p$-wave component). The latter is at the upper bound of the sensitivity of our present experiments.

\acknowledgments We thank Thomas Scaffidi for sharing his manuscript with us prior to publication, and Steve Kivelson for a number of helpful discussions and for commenting on our manuscript. A.C. is grateful for support from the Julian Schwinger Foundation for Physics Research. A.P. acknowledges support by the Alexander von Humboldt Foundation through the Feodor Lynen Fellowship. Work at Los Alamos was funded by Laboratory Directed Research and Development (LDRD) program, and A.P. acknowledges partial support through the LDRD. N. K. acknowledges the support from JSPS KAKNHI (Grant No. 18K04715). The work at UCLA was supported by the National Science Foundation, grant number DMR-1709304. A.C. and A.P. contributed equally.


%

\cleardoublepage

\setcounter{table}{0}
\setcounter{figure}{0}
\setcounter{equation}{0}
\renewcommand{\thefigure}{S\arabic{figure}}
\renewcommand{\thetable}{S\arabic{table}}
\renewcommand{\theequation}{S\arabic{equation}}

\onecolumngrid
\appendix
\section{Supplemental Material}

\begin{centering}
{\large \textbf{Evidence for even parity unconventional superconductivity in \sro}}

A. Chronister$^{1 \dagger\star}$, A. Pustogow$^{1 \dagger\star}$, N. Kikugawa$^{2}$, D. A. Sokolov$^{3}$, F. Jerzembeck$^3$, C. W. Hicks$^{3}$, A. P. Mackenzie$^{3,4}$, E. D. Bauer$^{5}$, S. E. Brown$^{1 \dagger}$

$^1$Department of Physics $\&$ Astronomy, UCLA, Los Angeles, CA 90095, USA;

$^2$National Institute for Materials Science, Tsukuba 305-0003, Japan;

$^3$Max Planck Institute for Chemical Physics of Solids, Dresden 01187, Germany;

$^4$Scottish Universities Physics Alliance, School of Physics and Astronomy, University of St Andrews, North Haugh, St Andrews KY16 9SS, UK;

$^5$Los Alamos National Laboratory, Los Alamos, New Mexico 87545, USA.

\end{centering}

\section{The discontinuous transition at $B_{c2}(T\to0)$}
While at low temperature there is good evidence that the thermal and magnetic responses both originate from field-induced quasiparticles, the thermodynamic discontinuities at the first-order transition at $B_{c2}(T)$ were previously explored in some detail~\cite{Yonezawa2013,Yonezawa2014,Amano2015}. These are thermodynamically constrained, expressed as the appropriate Clausius-Clapeyron relation, \textit{e.g.},
\begin{equation*}
\frac{dB_{c2}}{dT_c}=-\frac{\Delta S}{\Delta M}.
\end{equation*}
While $\Delta M$ is the total magnetization and includes diamagnetic shielding, it is dominated by the hyperfine part in \sro\ \cite{Amano2015,Murakawa2007}. The phase transition was found discontinuous for $T\lesssim0.8$ K, with specific heat~\cite{Yonezawa2014} and MagnetoCaloric Effect~\cite{Yonezawa2013} measurements extending to temperatures as low as 90 mK. At $T=$200 mK, the entropy jump is quoted as 10\% of the normal state value, $\Delta S=(0.1)\gamma_NT_c$, where $\gamma_n$= 37.5 mJ/mol-K$^2$. Also reported at 200 mK, $dB_{c2}/dT_c$= -0.2 T/K. These combined results lead to the expectation $\Delta M(200\textrm{mK})\simeq .6-.7\chi_{n}B_{c2}(T=200\textrm{mK}$.

Since in these cases the Zeeman energy is much larger than the thermal energy scale, the fractional entropy of the transition is not expected to be strongly temperature-dependent. In that case, the magnetization discontinuity inferred from the measurements reported here compare favorably. That is, the experiments indicate a slightly smaller discontinuity, $\Delta M_s\simeq0.4\chi_nB_{c2}$ with $\chi_{sc}\simeq0.9(10)^{-3}$ emu/mole.

\section{Comment on the orbital shift for $^{17}$O}

The $^{17}$O orbital shifts $K_o$ are of considerable importance here, since we use them to reference the Knight shifts $K_s(1_{\parallel},1_{\perp},2)$ relative to ``0''. Generally, $K_o$ can extend to greater than +1500 ppm in oxygen-containing molecules, that are dominated by the paramagnetic term, and are linked to an increased $2p$-bonding order~\cite{Figgis1962}. In this case, the $2p-\sigma$ non-bonding orbital is $\sim3/4$-filled, and the bonding $2p-\pi$ orbitals of most interest here, are $\sim7/8$ filled~\cite{Luo2019}. in Ref.~\cite{Ishida1998}, the orbital shifts $K_o(1_{\perp},2)$ were emprically found neglibibly small by way of a so-called $K-\chi$ plot, where temperature is an implicit parameter. In a similar fashion, $K_o(1_{\parallel})\simeq0.18\%$. These values were applied to the analysis leading to Fig. 3, although we used $K_o(2)=0.02\%$. More specifically, if $K_o(1_{\parallel})$ is taken as vanishingly small, the Knight shift $K_s(1_{\parallel})$, originating dominantly with the $p-\pi$ orbitals, acquires the wrong (unphysical) sign. See Fig. 3(a), where total shift is seen to change sign as the field-induced quasiparticle density is monotonically increased with the field strength.

The paramagnetic orbital shift for O($1_{\parallel}$) implies it follows from the specific unquenching of the angular momentum in the $p-\pi$ states. These states, in hybridizing with the Ru $t_{2g}$ orbitals form the bands $\alpha$, $\beta$, $\gamma$ crossing the Fermi surface. Thus, the perturbative methods applied somewhat successfully to the cuprate superconductors~\cite{Pennington1989} may be less useful here.

Note also that if we adopt $K_s(1_{\perp})>0$, a similarly unphysical field-dependent sign change is imposed on the corresponding hyperfine part, and the diamagnetic part is expected less than .02\%. \textit{This is the clearest constraint on setting the stated upper bound to the condensate fraction of the shift} to $<10\%$ of $K_{normal}$.

Finally, more relative uncertainty is associated with the O(2) site. $K_o$=0.0\% was assumed in Ref.~\cite{Imai1998}. Here, we took it as +0.02\%, also small on the scale of oxygen paramagnetic orbital shifts, and just 25\% compared to inferred normal state hyperfine (Knight) shift. Using this value, the results for O(2) match those for O($1_{\parallel},1_{\perp}$).

\section{Nuclear Hamiltonian Parameterization}
For the $^{17}$O nucleus with spin $I=\frac{5}{2}$, the nuclear spin hamiltonian consists of two parts:
\begin{align}
H=H_Q+H_z \\
H_z=\gamma \vec{I} \cdot (1+K) \cdot \vec{B} \\
H_Q = \frac{eQ}{2I(2I-1)\hbar}\vec{I}\cdot V \cdot \vec{I}
\end{align}

where $H_z$ is the Zeeman interaction and $H_Q $ is the nuclear quadrupole interaction. $K = K_s + K_{orb}$ is the total shift tensor, including both orbital and hyperfine contributions. $Q$ is the electric quadrupole moment of the nucleus and $V$ is the electric field gradient (EFG).

 The quadrupolar term has the general effect of splitting the degenerate Zeeman transitions, resulting in $2I$ resonance frequencies. Thus, in the case of \sro, which has three distinct oxygen sites under the application of in-plane field ($B || a$), we expect a full $^{17}$O spectrum of 15 lines.

By measuring these NMR lines, one can probe the electronic spin susceptibility $\chi$ of the material via the strength of the hyperfine interaction ($K_s$).  However, as Equations (1-3) imply, this effect must be differentiated from the other interactions contributing to the total Hamiltonian. If the parameters defining the nuclear quadrupole interaction are known, this can be readily done. Fortunately, the hyperfine shift, orbital shift, and EFG tensors are all independent of field in the normal state, making it possible to determine these parameters experimentally. So, by measuring all fifteen $^{17}$O resonances at various fields in the normal state ($B>B_{c2}$), one can overdetermine the normal state shift tensor $K_{norm}$ and the EFG tensor $V$. Then, since the quadrupolar and orbital shifts are independent of the superconducting phase transition, any discrepancy between the expected and measured resonance frequencies below $B_{c2}$ can be directly attributed to changes in $K_s$, hence the spin susceptibility.

The nuclear spin Hamiltonian is canonically expressed in the principle axes system (PAS) of $V$, using the convention for the diagonal entries $V_{zz}\geq V_{xx}\geq V_{yy}$. By doing this, the quadrupolar part can be written compactly as:
\begin{align}
 H_Q= \frac{\nu_Q}{6}[3I_z^2-\vec{I}^2+\eta(I_x^2-I_y^2)]
\end{align}

 where, $\nu_Q$ is the principle axis NQR frequency, proportional to $V_{zz}$ and $\eta$ is the asymmetry parameter given by $(V_{xx}-V_{yy})/V_{zz}$. Since the shift tensor is also diagonal in this frame for all three oxygen sites in \sro,  the Zeeman term can be expressed as
 \begin{align}
 H_z=\gamma [ B_x (1+K_{xx})I_x +B_y (1+K_{yy})I_y + B_z (1+K_{zz})I_z]
\end{align}
 where $\vec{B}$ is also written in the EFG basis.

With the Hamiltonian written in this form, there are 7 parameters to be determined for each oxygen site: $\nu_Q$, $\eta$, the three PAS components of $\textbf{K}$, and the two angles relating $\hat{B}$ to the EFG frame, $\theta$ and $\phi$. However, as explained below, both the angles determining $\hat{B}$ can be measured independently-- leaving only 5 parameter to be determined via normal state measurements.

 \section{Sample Alignment With Respect to Magnetic Field Direction}

 \begin{figure}[h]
\includegraphics[width=0.5\columnwidth]{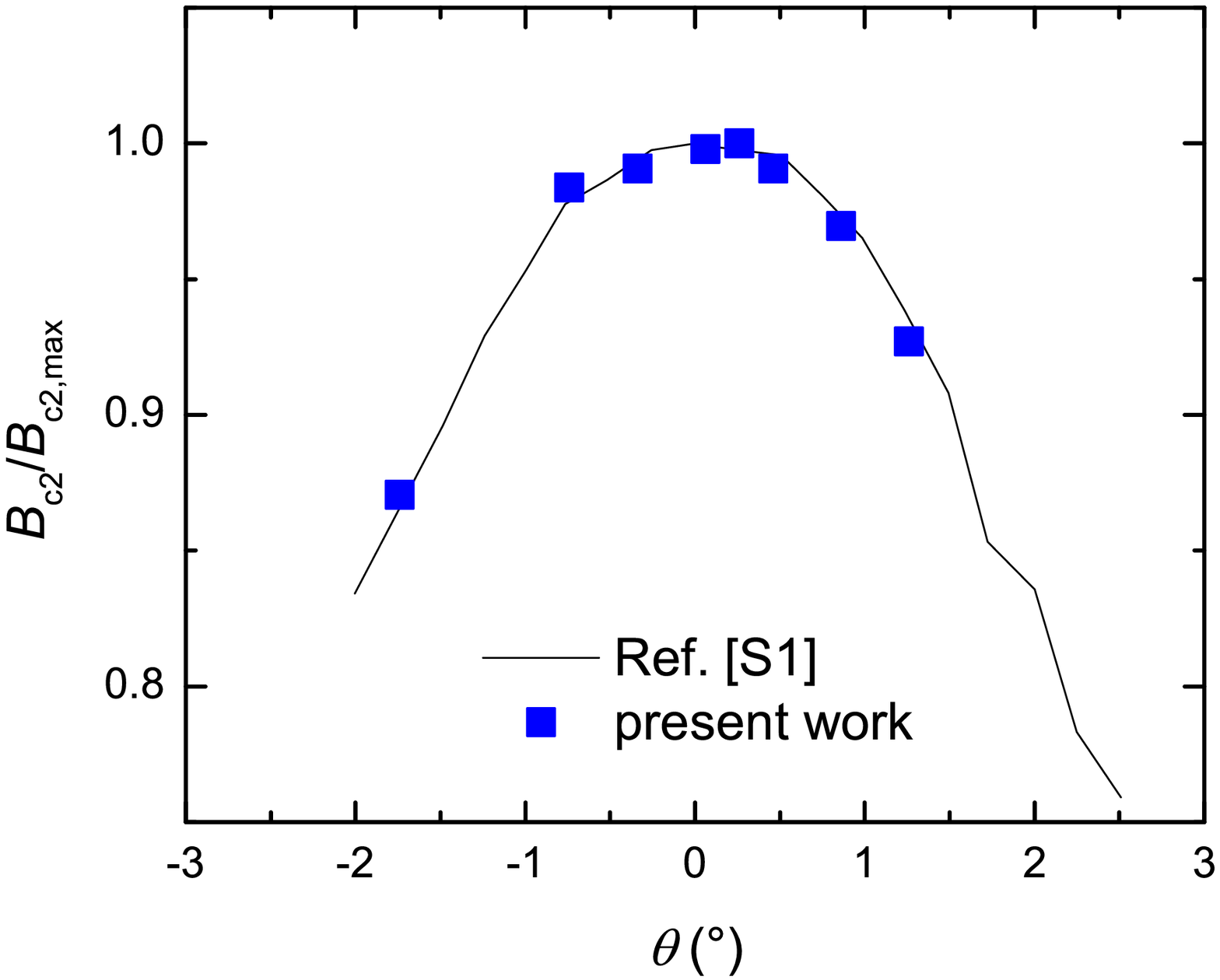}
\caption{The angle dependence of the upper critical field $B_{c2}$, determined from the field-dependence of the coil inductance~\cite{Pustogow2019}, was measured with a piezo-electric rotator (blue symbols). $B_{c2}(\theta)$ is plotted in units of the maximum value $B_{c2,max}= 1.42$~T. The data agree well with specific heat results from Ref.~\cite{Yonezawa2013} where $B_{c2,max}$ ranges from 1.41--1.45~T for different samples and field-sweep conditions. The in-plane condition is satisfied to $\pm 0.2^{\circ}$ for our sample.
}
\label{fig:Hc2_vs_angle}
\end{figure}

First, the out-of-plane angle can be determined independent of the NMR spectrum by utilizing the extreme anisotropy in the upper critical field for $B || ab$ and $B || c$. $B_{c2}$ reaches a maximum of around 1.45~T with the field aligned directly in plane \cite{Yonezawa2013}. As mentioned in the main text, the NMR coil containing the sample is mounted on a piezoelectric step rotator with rotation axis perpendicular to the applied field. By activating the piezo until $B_{c2}$ reaches a maximum, the in-plane condition can be aligned to within $\pm 0.2 \deg$. The angle dependence of $B_{c2}$ is shown in Fig.~\ref{fig:Hc2_vs_angle}.



The in-plane angle is then checked by \textit{a posteriori} visual inspection of the sample mounting using a microscope to view the sample orientation. While the in-plane condition was verified by anisotropy of $B_{c2}$ in Fig.~\ref{fig:Hc2_vs_angle}, a $3\deg$ angle is found between the long axis of the single crystal and the magnetic field direction. 

 \section{Normal-State Measurement of Hamiltonian Parameters}
 
\begin{figure}[]

\includegraphics[width=1\columnwidth]{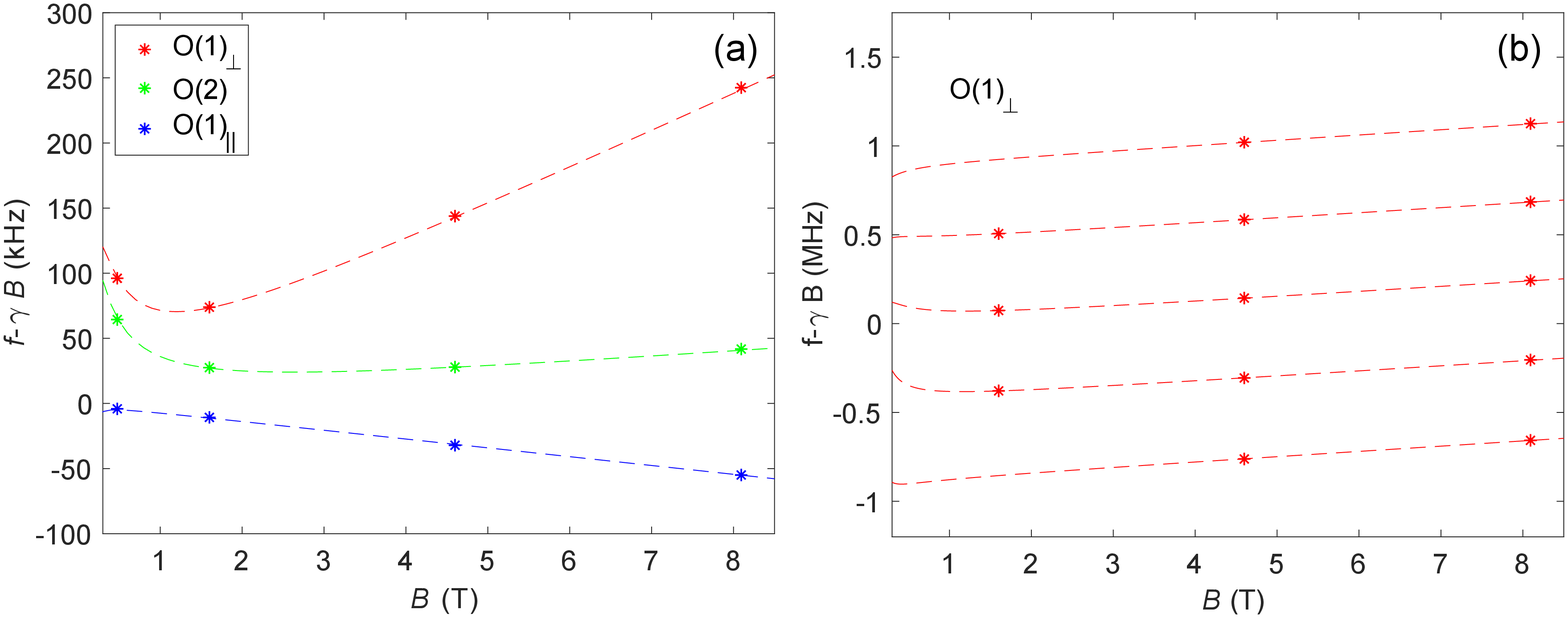}
\caption{Numerically calculated transition frequencies (dashed lines) compared to measured resonance frequencies (symbols) at different fields for (a) central transitions of the three oxygen sites and (b) central and satellite transitions of O(1$_{\perp}$).
}
    \label{calc_freqsweep_compare}
\end{figure}

\begin{table}[]
\begin{tabular}{ l | l | l | l  l}
            & Shift (\%) & NQR frequency (MHz) & Asymmetry &\\
            \hline
   O(1) &  &  &  &  \\
           & $K_{1||} = -0.12 $ &  $\nu_Q= 0.765$ &  $\eta = 0.174$ &  \\
           & $K_{1\perp} = +0.509  $&  &  & \\
           & & & & \\
           \hline
   O(2) &  &  &  &  \\
           & $K_{2ab} = +0.082 $  &  $\nu_Q= 0.6065$ & $\eta = 0$ & \\
           & & & &\\
           \hline

\end{tabular}

\caption{List of best fit Hamiltonian parameters for the different oxygen sites with $\theta = 0\deg$ and $\phi=3\deg$. The two planar oxygen sites O(1) and O(1') are identical without applied field and are labeled O(1). With the field aligned in the Ru-O plane ($\theta = 0\deg$), just two components of $K$ are relevant for the O(1) site while only one is relevant for the O(2) site. }
\label{table1}
\end{table}

The remaining parameters are determined by fitting the output of a numerical diagonalization of the exact Hamilton to experimentally measured $^{17}$O NMR transitions at three fields greater than $B_{c2}$, (B=1.6T, 4.6T, 8T). The quadrupolar parameters for \sro\ have been investigated on a different crystal in a previous study \cite{Luo2019}, and were used as a starting point for the fit. A comparison of the best fit calculation to the experimental normal-state line positions are shown in Fig.~\ref{calc_freqsweep_compare}. The fit reproduces the measured resonance frequencies extremely well, with an average error of less than 1kHz across the three fields. The parameters extracted from the best fit are given in Table \ref{table1}. These values are consistent with previously published results \cite{Mukuda1998}. 

\begin{figure}[]

\includegraphics[width=0.8\columnwidth]{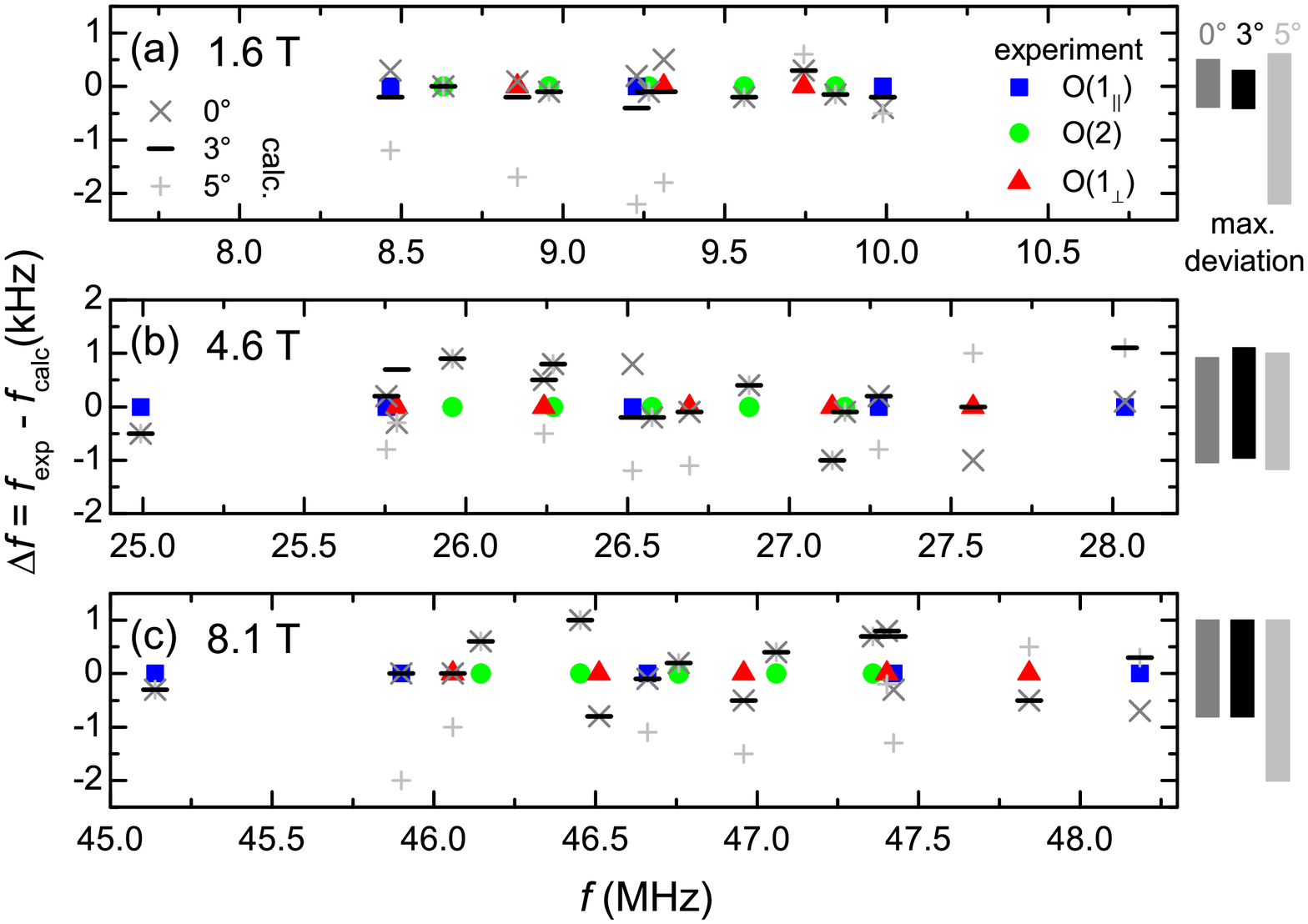}
\caption{Difference between calculated and measured resonance frequencies, $\Delta f=f_{exp}-f_{calc}(\phi)$ is shown for (a) $B=1.6$, (b) 4.6 and (c) 8.1~T. The experimental data ($\Delta f=0$) are shown at the respective frequency for O(1$_{\parallel}$), O(2), O(1$_{\perp}$) in blue, green and red colors, respectively. The calculated results at $\phi=0\deg$, $3\deg$ and $5\deg$ in-plane angle with respect to $\BB\parallel [100]$ are indicated by crosses, minus and plus signs in dark grey, black and light grey color, respectively.
On the right we illustrate the maximum deviations from the experimentally determined peak positions. The angle dependence becomes most pronounced at low fields, where $\phi=3\deg$ provides the best fit at $B=1.6$~T. The accumulated rms error is smallest for $\phi=3\deg$ at all fields.
}
    \label{calc_angle_compare}
\end{figure}

\section{Discussion of in-plane angle uncertainty}

Due to the weak dependence of the quadrupolar term on in-plane angle near $0\deg$  at high field, it is possible to accurately fit the normal-state spectra for a range of in-plane angles ($\approx 0\deg- 5\deg$). This is illustrated in Fig.~\ref{calc_angle_compare}, which shows the deviation between the predicted and measured frequencies for all $^{17}$O transitions using $0\deg$, $3\deg$, $5\deg$ in-plane angle fits; the predicted normal-state frequencies differ only by $\pm 1$~kHz between the fits for $B=1.6$--8~T. While the overall deviations are smallest for $\phi=3\deg$ (which was used for NMR shift analysis), any systematic error introduced by uncertainty in the in-plane angle should be examined.

\begin{figure}[h]
\includegraphics[width=1\columnwidth]{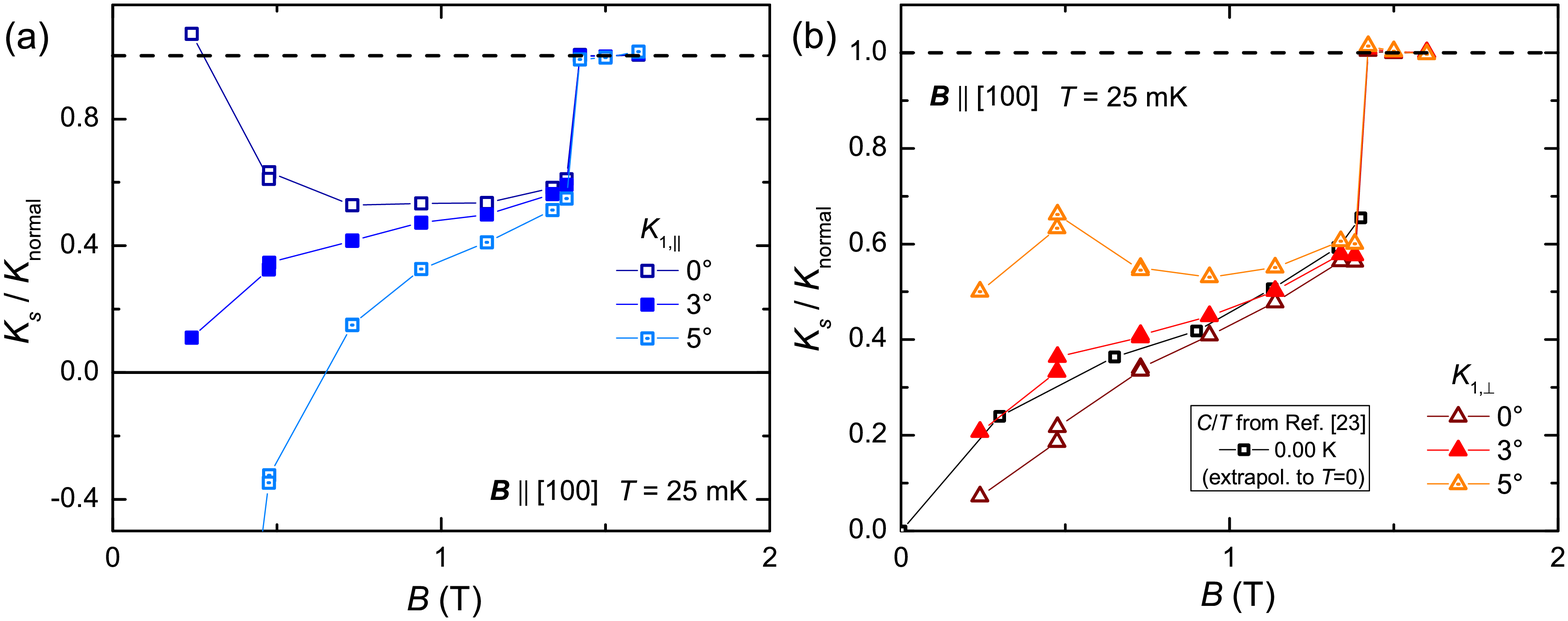}
\caption{Difference between extracted $K_s/K_{normal}$ for in-plane angles $\phi=0^{\circ}$, $3^{\circ}$ and 5$^{\circ}$ with respect to $\BB\parallel [100]$. (a) $\phi=0^{\circ}$ and 5$^{\circ}$ yield strong deviations for $K_{1\parallel}$ with non-physical behavior $K_s<0$ and $K_s>K_{normal}$. (b) Due to generally larger Knight shift, the variations of $K_{1\perp}$ are less pronounced, yielding more robust values. Still, the non-monotonous behavior upon lowering $B$ for 5$^{\circ}$ is not meaningful, and also the susceptibility values smaller (0$^{\circ}$) than the quasiparticle contribution from specific heat, $K_s/K_{normal}<C/C_{normal}$~\cite{NishiZaki2000}, are unphysical. Altogether, we conclude upon an in-plane angle $\phi=3^{\circ}$ from [100], consistently used in this work.
}
    \label{Kplot_angle_compare}
\end{figure}

While for fields $B>B_{c2}$ the effect of in-plane angle is small, it can have a strong impact on the expected normal-state position at lower fields. As such, this affects the ability to extract $K/K_{normal}$ for $B\rightarrow 0$. To illustrate this, the resulting $K/K_{normal}$ are shown for best fits using the three in-plane angles $\phi=0\deg$, $3\deg$ and $5\deg$ in Fig.~\ref{Kplot_angle_compare}. The  O(1$_{\parallel}$) site shows a particularly strong dependence on $\phi$: the  $0\deg$ and $5\deg$ fits produce unphysical behavior, with $K_s$ exceeding the normal-state value for $\phi=0\deg$ and changing sign for $\phi=5\deg$. The O(1$_{\perp}$) site has a much weaker dependence, but still shows unphysical behavior for angles deviating from $3\deg$. This gives further confidence in the visually determined $3\deg$ angle, but also shows that the O(1$_{\perp}$) site is more robust to a small angle systematic error in evaluating the Knight shifts. Additionally, it should be noted that the apical O(2) site, although having much weaker hyperfine coupling, is completely independent of the in-plane angle due to its axial symmetry, avoiding this issue altogether.

\end{document}